\documentclass[12pt,a4paper,floatfix]{iopart}
\usepackage{amssymb,graphicx,cite}
\begin{document}
 

 


 
 
   
 
\title[Expansion of a 
Bose-Einstein condensate on an optical lattice]{Expansion of
a
Bose-Einstein condensate formed on a joint harmonic and  one-dimensional
optical-lattice potentials}
 
\author{Sadhan K. Adhikari}
\address{Instituto de F\'{\i}sica Te\'orica, Universidade Estadual
Paulista, 01.405-900 S\~ao Paulo, S\~ao Paulo, Brazil}

\date{\today}

\begin{abstract}

We study the expansion of a Bose-Einstein condensate  trapped in
a combined optical-lattice and axially-symmetric harmonic potentials using
the numerical solution of the mean-field Gross-Pitaevskii equation. First,
we consider the expansion of such a condensate under the action of the
optical-lattice potential alone.  
In this case the result of numerical simulation for the axial and radial
sizes during expansion is in agreement with two  experiments by
Morsch {\it et al.} [2002 {\it Phys. Rev. A} {\bf 66} 021601(R) and 2003
{\it Laser Phys. } {\bf 13} 594].  Finally, we consider the expansion
under the action of the harmonic potential alone.  In this case the
oscillation and the disappearance and revival of the resultant
interference pattern is in agreement with the experiment by M\"uller {\it
et al.} [2003 {\JOB} {\bf 5} 538].

\end{abstract}

 

\section{Introduction}

The experimental loading of a cigar-shaped Bose-Einstein condensate (BEC)
in both one- \cite{1,2} and three-dimensional \cite{greiner} optical
lattice potentials generated by standing-wave laser fields has initiated a
new class of investigations.   The optical-lattice potential has been used
in the study of matter-wave interference \cite{1}, of oscillating atomic
current in a one dimensional array of Josephson junctions \cite{cata,ady},
of
Bloch oscillation and Landau Zenner tunneling \cite{ari0},
of
superfluid-insulator classical \cite{cata2} and quantum \cite{greiner}
phase transitions, and  in the generation of matter-wave bright soliton
\cite{sol}.  


As the size of the condensates in above experiments is much too small, one
may resort to an expansion in order to photograph, observe or study
it.  Actually, expansion of the condensate formed on an optical
lattice was crucial in the confirmation of Josephson oscillation across an
array of one-dimensional junctions \cite{cata}, of Bloch oscillation and
Landau-Zenner tunneling \cite{ari0}, of superfluid to Mott insulator
quantum phase transition \cite{greiner}, and of transition from 
superfluid to a classical insulator \cite{cata2,th}. 
Hence it is of interest
to
perform a theoretical study of the  expansion of the BEC formed on a joint
one-dimensional optical-lattice and axially-symmetric harmonic traps upon
the removal of either  trap and compare with available
experiments \cite{ari1,ari2,ari3}.  Here we undertake this study using the
mean-field
axially-symmetric Gross-Pitaevskii (GP) equation \cite{8}. There have been
similar
studies of free expansion of an axially-symmetric BEC in the absence of an
optical-lattice potential \cite{fexp}.

For the present study we consider the expansion of a BEC on a joint
one-dimensional
optical-lattice and axially-symmetric harmonic potentials with the
optical-lattice  aligned along the axial direction of the
harmonic potential. First we consider switching off the harmonic trap
alone. In this case the condensate expands along the radial direction
maintaining an almost  constant transverse size. We also reconsider
the  free expansion of such a  BEC formed in the absence of
an optical-lattice potential. In both cases the time evolution of the
$e^{-1}$ half-width of a Gaussian fit to the density profile in the axial
and radial directions are obtained and found to be in good agreement with
experiments by Morsch {\it et al.} \cite{ari1,ari2}. Next we consider
switching off the optical-lattice trap alone. The periodic
one-dimensional optical-lattice potential divides the whole BEC in
parallel slices. Upon the removal of the   optical-lattice trap these
slices expand and interfere with each other. The result is the formation
of an interference pattern of three peaks. 
However, in the presence of the harmonic trap the side peaks execute
simple-harmonic oscillation with a definite frequency and
amplitude. If the interference pattern is allowed to evolve in the
harmonic trap for some hold time and the harmonic trap is then switched
off, prominent interference pattern continues  for small hold time. With
the increase of hold time the interference pattern disappears and revives
periodically.  In this case the results of the simulation is in agreement
with experiment of M\"uller {\it et al.} \cite{ari3}.  

In section 2 we present our mean-field model based on the GP equation. 
In section 3 we present the numerical results and a comparison with
available experiments. Finally, the conclusions are given in section 4.

\section{Mean-Field Model}

We base the present study on the numerical solution of the
time-dependent GP equation \cite{8} in the presence
of a
combined axially-symmetric
harmonic and optical-potential traps. 
The time-dependent BEC wave
function $\Psi({\bf r};t)$ at position ${\bf r}$ and time $t $
is described by the following  mean-field nonlinear GP equation
\cite{8}
\begin{eqnarray}\label{a} \left[- i\hbar\frac{\partial
}{\partial t}
-\frac{\hbar^2\nabla^2   }{2m}
+ V({\bf r})
+ gN|\Psi({\bf
r};t)|^2
 \right]\Psi({\bf r};t)=0,
\end{eqnarray}
where $m$
is
the mass and  $N$ the number of atoms in the
condensate, $\hbar=h/(2\pi)$ with $h$ the Planck's constant, 
 $g=4\pi \hbar^2 a/m $ the strength of atomic interaction, 
$a$ the  scattering length.  In the presence of the combined
harmonic  and optical traps 
     $  V({\bf
r}) =\frac{1}{2}m \omega ^2(\rho ^2+\nu^2 z^2) +V_{\mbox{opt}}$ where
 $\omega$ is the angular frequency of the harmonic trap 
in the radial direction $\rho$,
$\omega_z \equiv \nu \omega$ that in  the
axial direction $z$, and $V_{\mbox{opt}}$ is
the optical-lattice trap introduced later.  
The normalization condition  is
$ \int d{\bf r} |\Psi({\bf r};t)|^2 = 1. $

In the axially-symmetric configuration, the wave function
can be written as 
$\Psi({\bf r}, t)= \psi(\rho ,z,t)$.
Now  transforming to
dimensionless variables $\hat \rho =\sqrt 2 \rho /l$,  $\hat z=\sqrt 2
z/l$,
$\tau=t
\omega, $
$l\equiv \sqrt {\hbar/(m\omega)}$,
and
${ \varphi(\hat \rho,\hat z;\tau)} \equiv   \hat \rho \sqrt{{l^3}/{\sqrt
8}}\psi(\rho ,z;t),$   (\ref{a}) becomes \cite{11}
\begin{eqnarray}\label{d1}
&\biggr[&-i\frac{\partial
}{\partial \tau} -\frac{\partial^2}{\partial
\hat \rho^2}+\frac{1}{\hat \rho}\frac{\partial}{\partial \hat \rho}
-\frac{\partial^2}{\partial
\hat z^2}
+\frac{1}{4}\left(\hat \rho^2+\nu^2 \hat z^2\right) \nonumber \\
&+&\frac{V_{\mbox{opt}}}{\hbar \omega} -{1\over \hat \rho^2}  +                                                          
8\sqrt 2 \pi n\left|\frac {\varphi({\hat \rho,\hat z};\tau)}{\hat
\rho}\right|^2
 \biggr]\varphi({ \hat \rho,\hat z};\tau)=0, 
\end{eqnarray}
where nonlinearity
$ n =   N a /l$. In terms of the 
one-dimensional probability 
\begin{equation}\label{pzt}
 P(z,t) \equiv 2\pi$ $\int_0 ^\infty 
d\hat \rho |\varphi(\hat \rho,\hat z,\tau)|^2/\hat \rho , 
\end{equation}
the
normalization
of
the
wave
function 
is given by $\int_{-\infty}^\infty d\hat z P(z,t) = 1.$  
The probability 
$P(z,t)$ is  useful in the study of the present problem under the
action of the optical-lattice potential, specially in the
investigation of the formation and
evolution of the interference pattern after the removal of the
trapping potential(s).

The  mean-field GP equation (\ref{a}) can be  derived for a dilute and
weakly-interacting system.
This is based on the
condition that the scattering length $a$ be much smaller than the average
distance between atoms and $N>>1$. Hence, the 
dilute gas approximation can be written as $\bar n |a|^3<<1$
\cite{8}, where $\bar n$ is the average density of the gas. 
However, for a simple and practical estimate of the weakly interacting
condition one should compare the interaction energy $E_{\mbox{int}}$ with
the
kinetic energy of the system $E_{\mbox{kin}}   $. From theoretical
consideration it follows that \cite{8}
\begin{equation}
\frac{E_{\mbox{int}}   }{E_{\mbox{kin}}   } \propto \frac{N|a|}{l}.
\end{equation}
The GP equation is valid for small $N|a|/l$. Systematic previous
studies have shown this equation to be valid for  $N|a|/l$ as large as
several hundreds \cite{8} and the domain of the experiments of
\cite{ari1,ari2,ari3} is well within this limit.

In the experiments of \cite{ari1,ari2,ari3} a completely asymmetric trap
is employed with trapping frequencies in three perpendicular directions 
$x$, $y$ and $z$ taken  in the ratio 2:1:$\sqrt 2$. An analysis
of the experiments using such a fully  asymmetric trap is beyond present 
numerical possibilities. However, in the experiment only a section of the
condensate is observed in one of the $x$-$y$, $y$-$z$, or $z$-$x$ planes,
e.g., the $x$-$z$ plane. The
optical-lattice potential is set in one of the in-plane directions, e.g., 
the $z$ direction. To make the problem numerically tractable, the trapping
frequency in the  perpendicular $y$ direction will be  set equal to that
in the $x$ direction. This seems to be reasonable as the anisotropy in
the experiment is not too large.

We shall consider several aspects of free expansion of a BEC formed on
a horizontal  optical-lattice plus harmonic potentials 
in the experiments  by  Morsch  {\it et
al.}. \cite{ari1,ari2} and  M\"uller  {\it et al.} \cite{ari3}.
In these experiments with repulsive $^{87}$Rb atoms
the imaging is done in a certain direction and the
observed trapping frequencies are $\omega_x= 2\pi\times 25$ Hz in the
radial
direction ($\rho$) and  $\omega_z=\omega\nu= 2\pi \times 35.5$ Hz in the
axial optical-lattice 
($z$) direction. The frequency in the unobserved perpendicular direction
$y$ is approximated by that in the $x$ direction. The actual experimental
frequency in the $y$ direction is $\omega_y= 2\pi\times 25/\sqrt 2$ Hz.
The
optical
potential created with the standing-wave laser field 
is given by $V_{\mbox{opt}}=V_0E_R\sin^2 (\pi z/d)$,
with $E_R=\hbar^2\pi^2/(2md^2)$,   $V_0$  the 
strength, and $d$ the lattice spacing. For the mass $m=1.441\times
10^{-25}$ kg of $^{87}$Rb the
harmonic
oscillator length $l=\sqrt {\hbar/(m\omega)} = 2.160$ $\mu$m 
and the 
present
dimensionless time unit  $\omega ^{-1} =
1/(2\pi\times 25)$ s $= 6.37$ ms. In terms of the dimensionless spacing 
 $d _0= \sqrt2 d/l $ and  the dimensionless 
standing-wave energy parameter $E_R/(\hbar \omega)= \pi^2/d _0^2$,
 $V_{\mbox{opt}}$ of 
  (\ref{d1}) is
\begin{equation}\label{pot}
\frac{ V_{\mbox{opt}}}{\hbar \omega}=V_0\frac{\pi^2}{d_0^2} 
\left[
\sin^2 \left(
\frac{\pi}{d_0}\hat z
\right)
 \right].
\end{equation}
Although we employ the dimensionless space units $\hat \rho$ and $\hat z$
and
time unit
$\tau$ in numerical calculation, the results are reported in actual units 
$r$ $\mu$m, $z$ $\mu$m  and $t$ ms and compared with the experiments
of  
Morsch  {\it et al.} \cite{ari1,ari2} and  M\"uller  {\it et al.}
\cite{ari3}.

\section{Numerical Results}
 
We solve   (\ref{d1}) numerically  using a   
split-step time-iteration
method
with  the Crank-Nicholson discretization scheme described recently
\cite{11x}.  
The time iteration is started with the  harmonic oscillator solution
of   (\ref{d1}) with
 $n=0$: $\varphi(\hat \rho,\hat z) = [\nu
/(8\pi^3)  ]^{1/4}$
$\hat \rho e^{-(\hat \rho^2+\nu \hat z ^2)/4}$ 
\cite{11}. First
the
nonlinearity $n$  is  
slowly increased by equal amounts in $10000n$ steps
of 
time iteration until the desired value of $n$ is
attained. Then the optical-lattice strength $V_0$ is increased in 10 000
small steps and the time iteration continued until the final value of 
 $V_0$ is attained. 
Then, without changing any
parameter, the solution so obtained is iterated 50 000 times so that a
stable
solution  is obtained 
independent of the initial input
and time and space steps.

The   BEC  on the
optical
lattice  for a specific nonlinearity and the 
interference pattern upon the  free expansion 
of such a BEC have been
recently studied using the numerical solution of
the GP equation  while both the optical-lattice and harmonic potentials
are
switched off in one, two and three dimensions \cite{ady,adhi1}. Here we
study
the
expansion of the BEC when
either the harmonic potential or the optical-lattice potential is switched
off. The periodic optical-lattice potential cuts the condensate into
several pieces in different sites.  As the present calculation is
performed with the full wave function without approximation, phase
coherence among different optical-lattice sites is automatically
guaranteed in the initial state. As a result when the condensate is
released from the periodic optical-lattice trap, a matter-wave
interference pattern is formed in a few milliseconds.  The atom cloud
released from one lattice site expand, and overlap and interfere with atom
clouds from neighboring sites to form the robust interference pattern.  
consisting  of a central peak and two smaller symmetrically spaced peaks 
moving in opposite directions.  If the harmonic
potential is kept on while switching off the optical-lattice potential the
two side peaks execute simple harmonic oscillation with the axial
frequency $\omega_z$. This simple harmonic motion can be explained
considering the dynamical evolution of the interference peaks in the
harmonic potential \cite{ari3,xiong}. Since the lattice transfers momentum
to
the condensate in units of $2p_R= h/d$, the
recoil velocity is given by $2v_R=2p_R/m= h/(md)$ and the two side peaks
move with velocities $\pm h/(md)$ at time $t=0$ immediately after release.
If the harmonic trap is maintained the two peaks will oscillate with an
amplitude $h/(md\omega_z)$. Consequently, the motion of the side peaks is
given by \cite{ari3}
\begin{equation}\label{shm} z(t)=\pm
\frac{h}{md\omega_z}\sin{(\omega_z t)}.
 \end{equation}

The experiments of Morsch {\it et al.} \cite{ari1,ari2} and 
M\"uller {\it
et al.} \cite{ari3} were
performed with about $(1-3)\times 10^4$ Rb atoms with the optical
lattice spacing $d=1.56$ $\mu$m. In our simulation we use
$N=20000$ atoms, $\omega=2\pi \times 25$ Hz and $\omega_z= 2\pi \times
35.5$ Hz. With the scattering length $a=5.77$ nm, this corresponds to a
nonlinearity of $n=Na/l\approx 53.4$. First we consider the results of
free expansion reported in figures 1 and 2 of \cite{ari1} and figure 1 of
\cite{ari2}. These results are related to the expansion of the condensate
when only the harmonic potential is switched off. Later we consider the
results reported in \cite{ari3} about the simple harmonic oscillation of
the interference peaks of the condensate when only the optical-lattice
potential is switched off as well as the disappearance and revival of 
interference pattern after a time evolution in the harmonic trap. 

\begin{figure}
 
\begin{center}

\includegraphics[width=.45\linewidth]{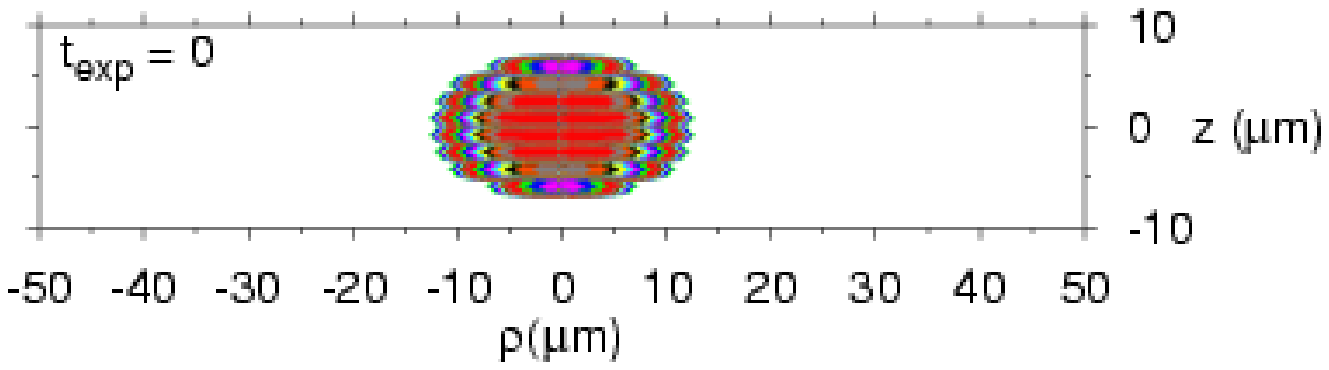}
\includegraphics[width=.45\linewidth]{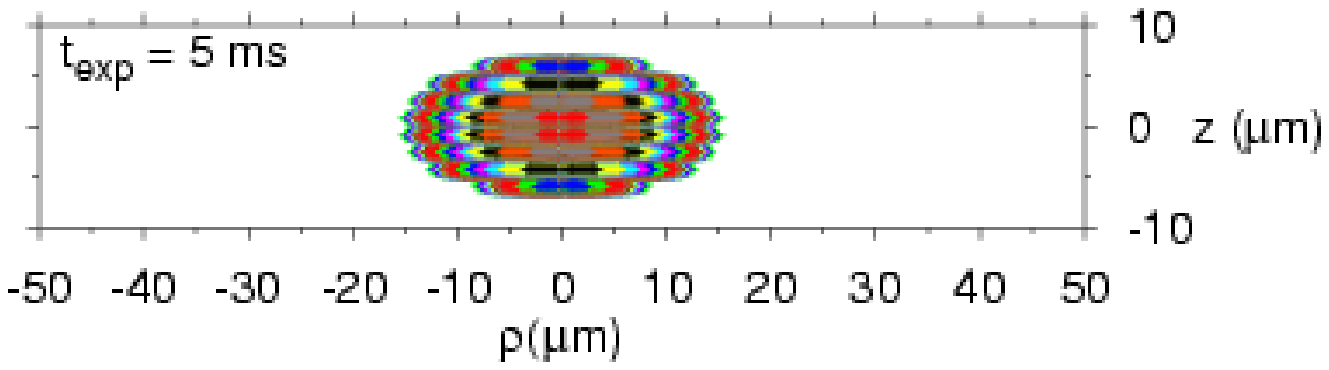}
\includegraphics[width=.45\linewidth]{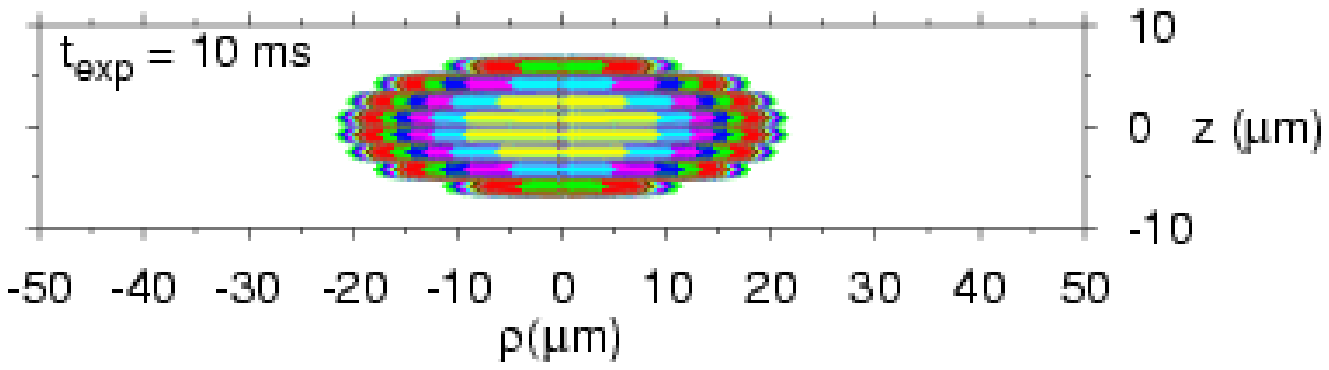}
\includegraphics[width=.45\linewidth]{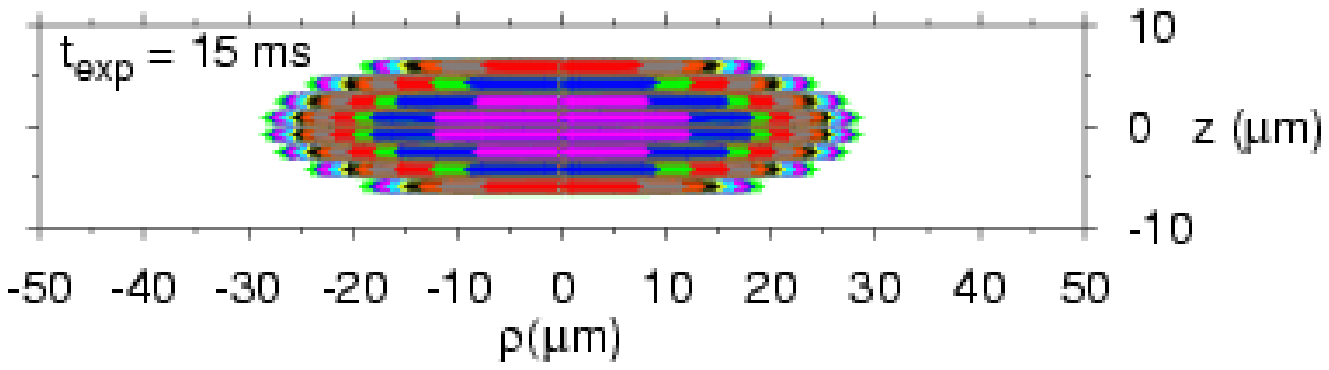}
\includegraphics[width=.45\linewidth]{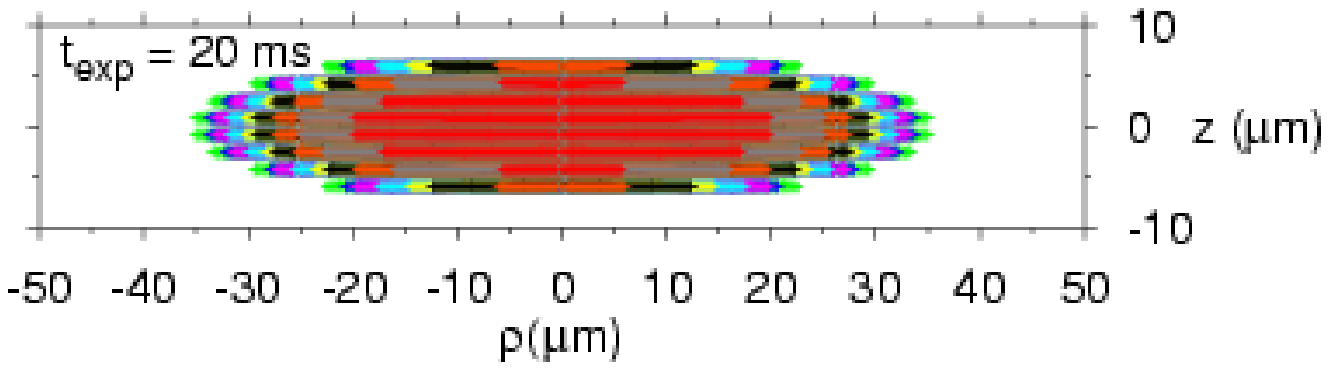}
\includegraphics[width=.45\linewidth]{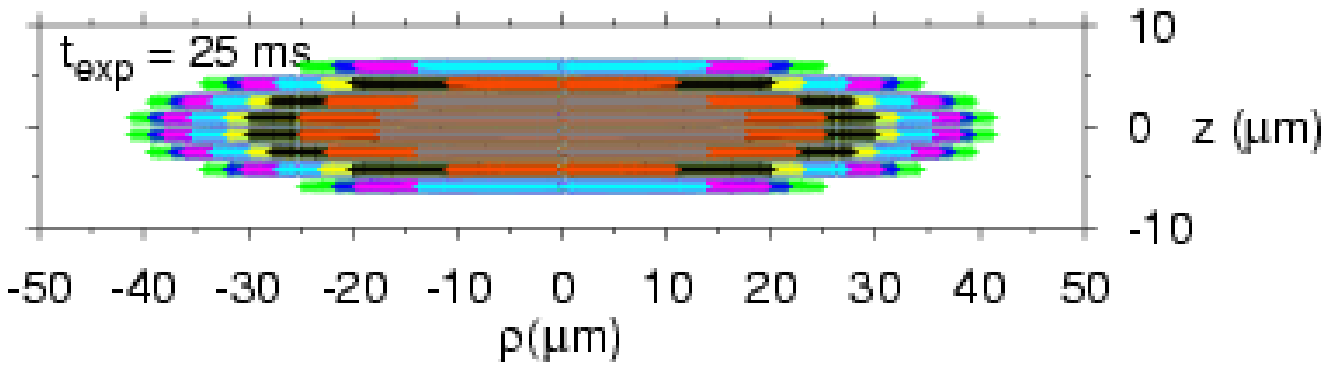}
\end{center}
 
\caption{View of the condensate in the radial direction after an expansion
time $t_{\mbox{\mbox{exp}}} = 0,$ 5 ms, 10 ms, 15 ms, 20 ms and 25 ms  in
the
optical-lattice
trap
alone obtained from the contour
plot of the wave function $|\psi(\rho,z,t_{\mbox{exp}})|$ for
$V_0=25,$
$n=53$, $\omega= 2\pi \times 25$ Hz and $\omega_z= 2\pi \times 35.5$
Hz. To obtain the full
section we used $|\psi(-\rho,z,t_{\mbox{exp}})|=
|\psi(\rho,z,t_{\mbox{exp}})|$.    
} \end{figure}

\begin{figure}[!ht]

\begin{center}

\includegraphics[width=.48\linewidth]{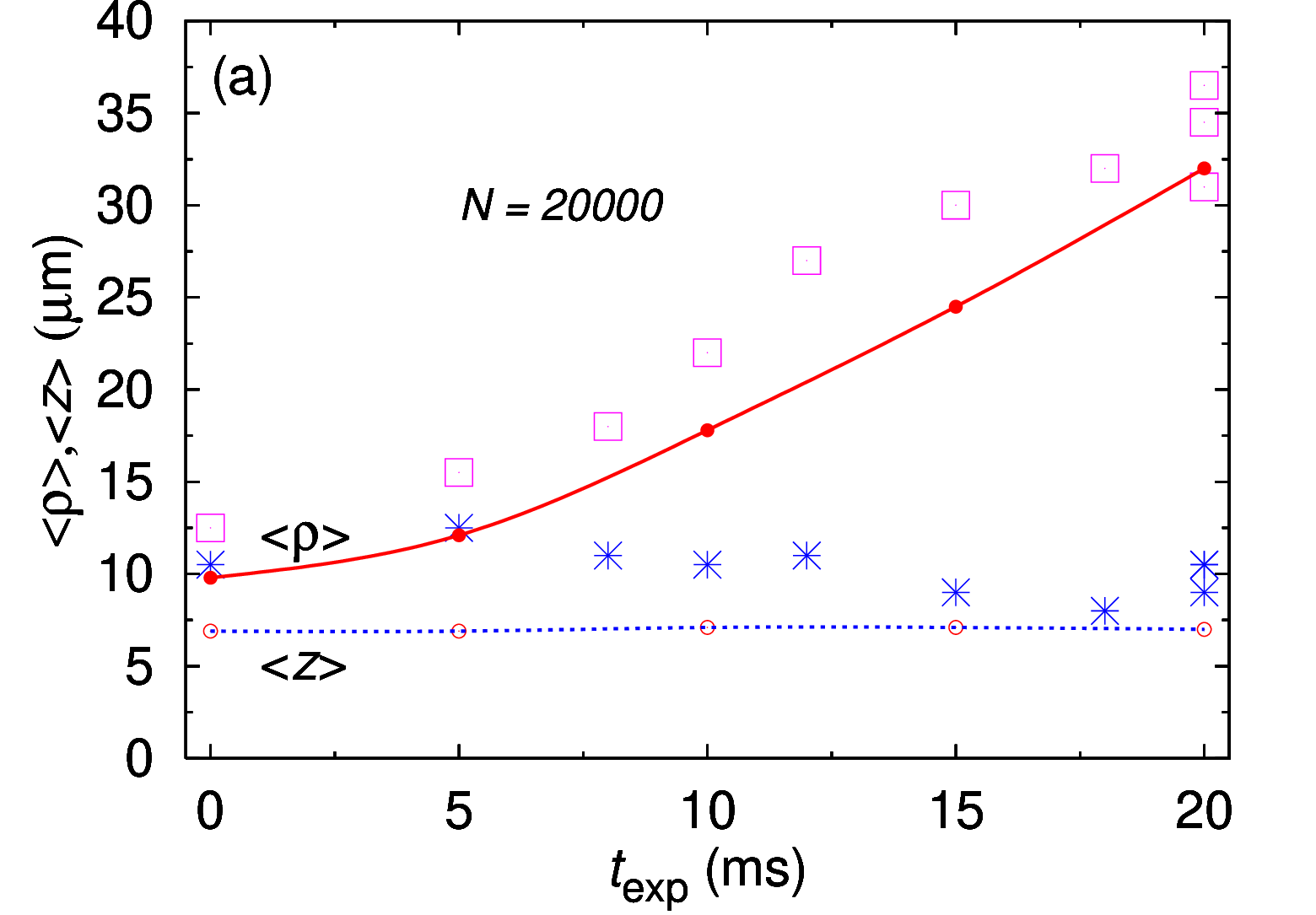}
\includegraphics[width=.48\linewidth]{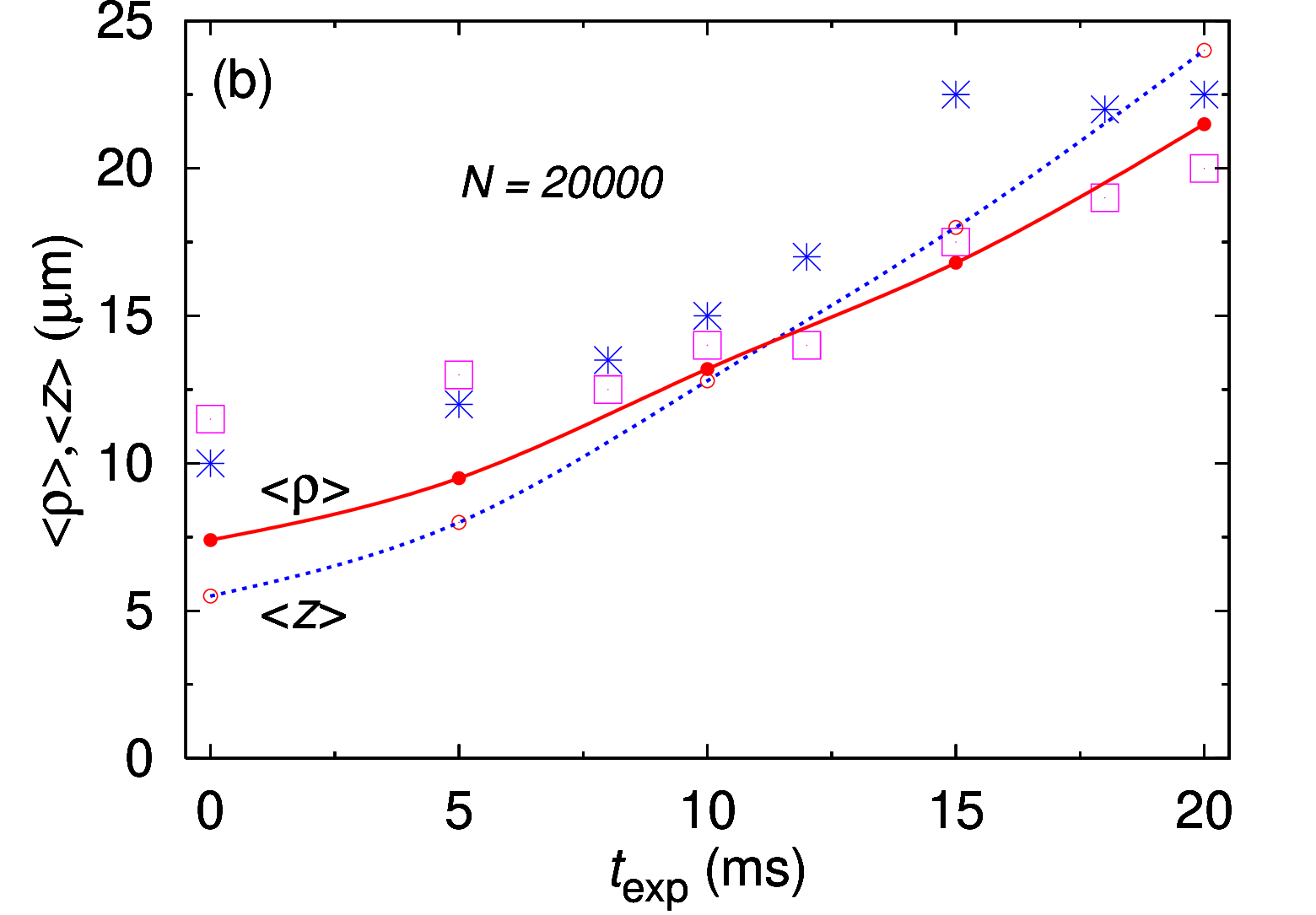}
\end{center}
 
\caption{Condensate dimensions $\langle \rho \rangle$ (radial) and
$\langle z
\rangle$ (axial) versus
the expansion
time $t_{\mbox{exp}}$ in the presence of   a lattice of strength
(a) $V_0=25$ 
and (b) $V_0=0$. The nonlinearity and frequencies are the same as in
figure 1. Open square: $\langle \rho \rangle $ of the experiment of
\cite{ari1}; star:
$\langle z\rangle $ of  the experiment of \cite{ari1}; full
line:  $\langle\rho
\rangle $
of the
present simulation; dotted line: $\langle z\rangle$  of the
present simulation.}  
 \end{figure}

\begin{figure}[!ht]

\begin{center}

\includegraphics[width=.5\linewidth]{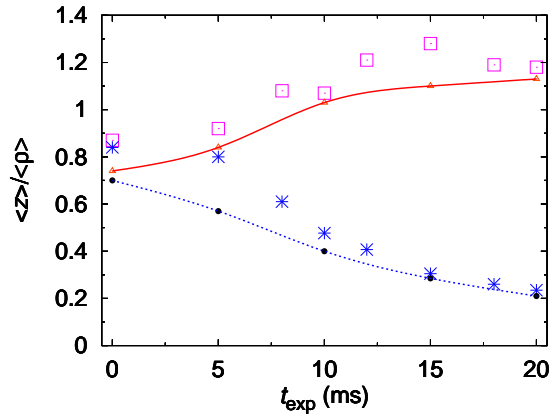}
\end{center}

\caption{The ratio  $ \langle z \rangle/ \langle\rho \rangle$ 
versus the expansion
time $t_{\mbox{exp}}$ in the presence of    lattices of strength
$V_0=20$ 
and  $V_0=0$. The nonlinearity and frequencies are the same as in
figure 1. Open square: experiment of
\cite{ari2} for $V_0=0$; star:
experiment of \cite{ari2} for $V_0=20$; dotted line: present simulation
for $V_0=0$; 
full line: 
present simulation  for $V_0=20$.}

 \end{figure}

To investigate the expansion of the condensate when the harmonic trap
alone is
switched off we use  optical-lattice strength $V_0=
25$. Once the
condensate is prepared on the joint optical-lattice and harmonic
potentials the harmonic potential is switched off and the expansion
studied under the influence of the optical trap alone. The view of the
condensate at different expansion times $t_{\mbox{exp}}=0$, 5 ms, 10 ms,
15 ms,  20 ms
and 25
ms are shown in figure 1.  The view at $(20-25)$ ms  is similar to figure
1 (b) of \cite{ari1} after 23 ms. The strong optical lattice divides the
condensate into smaller pieces trapped in different lattice sites. With
time one piece of BEC in a particular site   expand perpendicular to the
lattice without ever moving to a neighboring lattice. Consequently, the
BEC expands only in the radial direction and  maintains a constant size
in the axial direction. 

Next we calculate $\langle z \rangle$ and  $\langle \rho \rangle$ 
which are the $e^{-1}$ half-widths of Gaussian fits to the density profile
in the lattice (axial) and  radial directions, respectively, for different
expansion times. The results are plotted in figure 2 (a) for $V_0=25$ and
compared
with the experiment of \cite{ari1}. In figure 2 (b) we plot the same
results of expansion as in figure 2 (a) for $V_0=0$ and compare with the
experiment \cite{ari1}. Considering the 5 $\mu$m resolution of the imaging
system \cite{ari2} the agreement between the present numerical simulation
and experiment is good.

\begin{figure}

\begin{center}

\includegraphics[width=.48\linewidth]{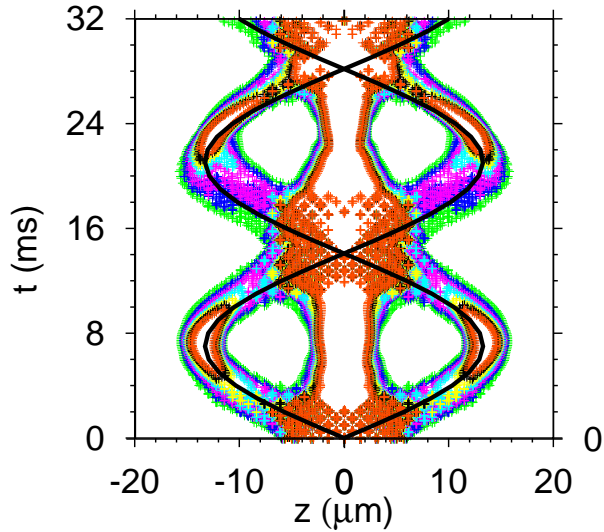}
\end{center}
 
\caption{Contour density plot of one-dimensional probability  $P(z,t)$ of
(\ref{pzt})
versus $z$ and $t$ for the initial BEC of figure 1 after the removal of
the
optical trap at $t=0$ when an interference pattern of a central peak and
two side peaks is formed. The sinusoidal pattern of  oscillation 
appears in the contour density plot and is compared with the simple
theoretical prediction (\ref{shm}), e.g., $z(t)=\pm 13.3 \sin (2\pi\times
35.5 
t)$ $\mu$m, shown by the
dark solid line.}
\end{figure}

\begin{figure}
 
\begin{center}

\includegraphics[width=.42\linewidth]{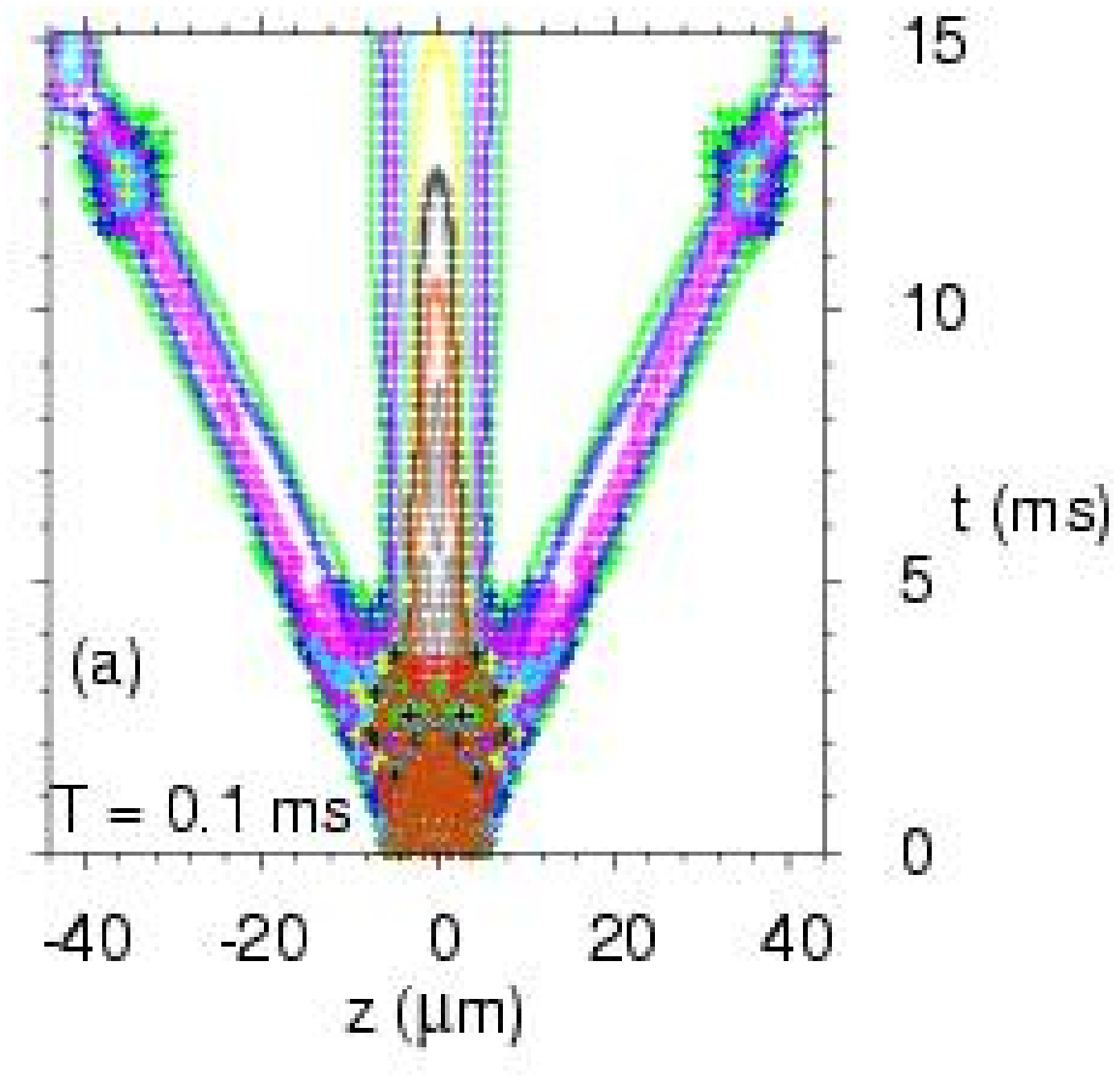}
\includegraphics[width=.42\linewidth]{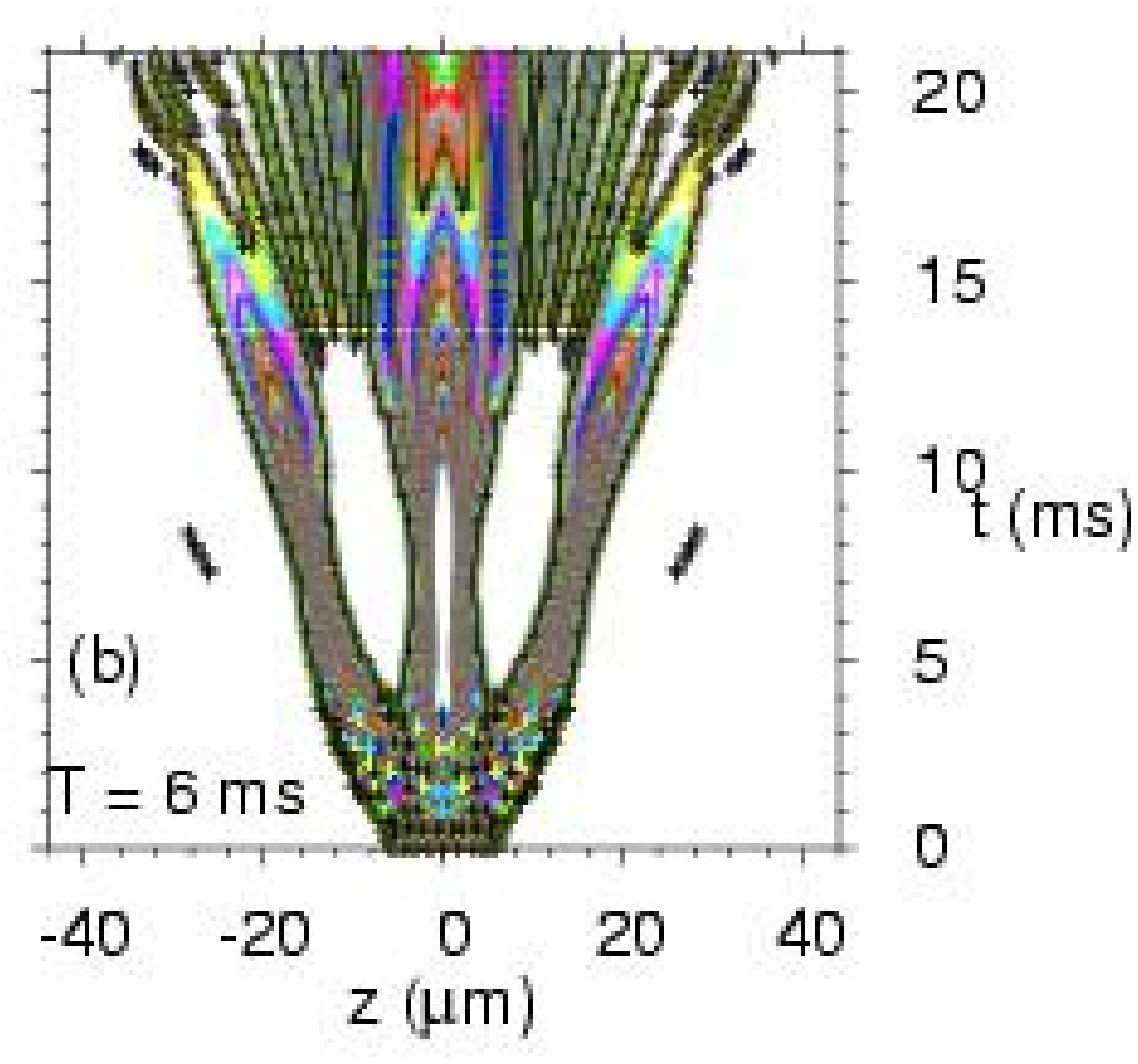}
\end{center} 

\begin{center}      

\includegraphics[width=.42\linewidth]{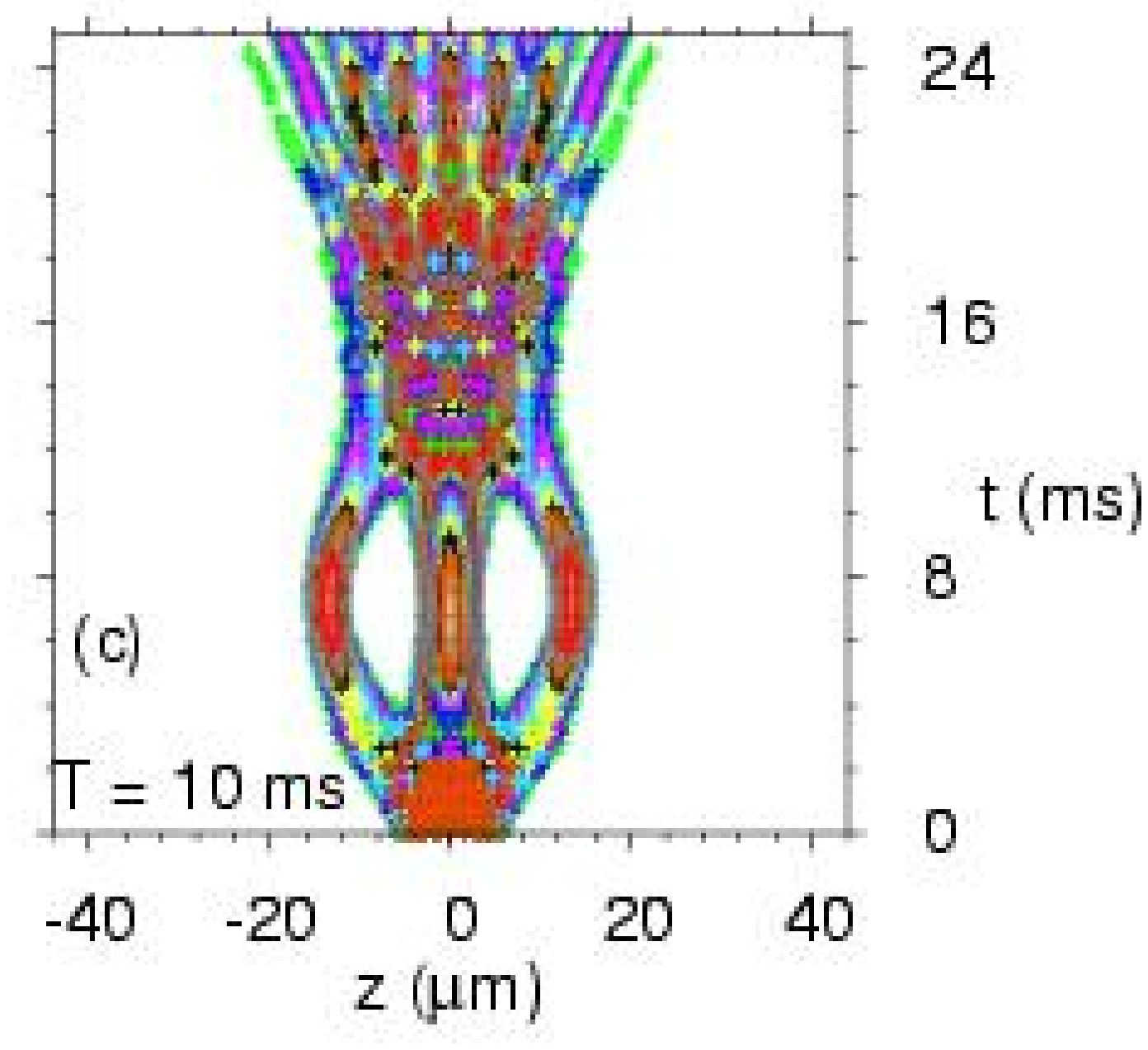}
\includegraphics[width=.42\linewidth]{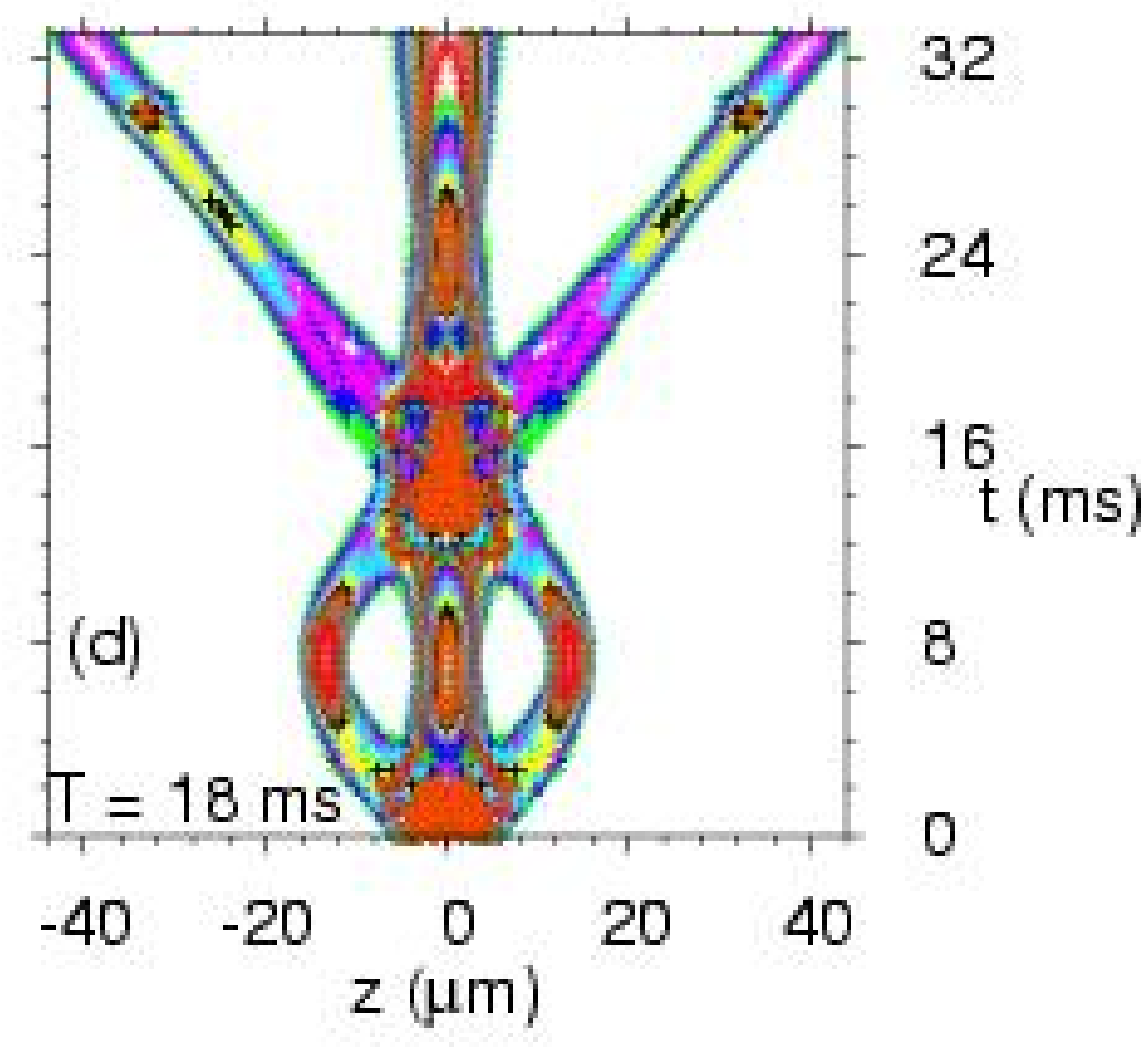}
\end{center}

\caption{Contour density plot of one-dimensional probability  $P(z,t)$ of
(\ref{pzt})
versus $z$ and $t$ for the initial BEC of figure 1 after the removal of
the
optical trap at $t=0$
 when an interference pattern of a central peak and
two side peaks is formed. The BEC is then kept in the harmonic trap for
hold times 
(a) $T=0.1$ ms,
(b) $T=6$ ms,
(c) $T=10$ ms and
(d) $T=18$ ms before finally removing it. The BEC is then
allowed to expand for 15 ms in order to see the interference peaks. The
interference pattern of three peaks represented by three trails  appears
for small hold times (a), becomes blurred (b) and disappears (c) with the
increase of hold time and
finally reappears  (d) when the hold time is larger than the half period
of
oscillation of the peaks given by (\ref{shm}).  }

 \end{figure}

Next we perform the same expansions of figure 2 with  optical lattices 
of strength $V_0=20$ and 0 and plot the ratio $\langle z \rangle/
\langle \rho \rangle$ in these cases for different expansion times and
compare with the experimental results of \cite{ari2}. The agreement is
good. In both figures 2
and 3 the agreement  is better for large expansion times. At large
expansion times the size is larger and hence possibly has less
experimental error  thus leading to a better agreement with experiment.

M\"uller {\it et al.} \cite{ari3} performed a different type of expansion
experiment on
a BEC
formed on a joint optical-lattice and harmonic traps upon switching off
only the optical-lattice trap. The condensate pieces from different sites
expand, overlap and interfere to form a pattern with three peaks which
oscillate according to (\ref{shm}). In figure 4 we show the contour
density plot of $P(z,t)$ of (\ref{pzt}) as functions of  $z$ and $t$. This  
clearly shows the simple harmonic motion of the two side peaks which is
compared with the simple theoretical prediction (\ref{shm}). 
The agreement
between the two is good.

Finally, M\"uller {\it et al.} \cite{ari3} observed in a remarkable
experiment the evolution of the interference patterns (formed upon
switching off the optical-lattice trap)
in the harmonic trap alone for different times. Then they removed the
harmonic trap and observed the free expansion of the BEC. For small hold
times $T$ in the magnetic trap the interference pattern is clearly
visible. They found that with the increase of hold time the prominent
interference pattern first becomes blurred and then 
disappears completely. Then it
reappears prominently
after a complete half cycle of oscillation according to (\ref{shm}). In
this case the period of oscillation is  28.2 ms. The
numerical simulation at $t=0.1$ ms in figure 5 (a) shows the clear trails
corresponding to the 
interference pattern
in agreement with  \cite{ari3}. The complete free expansion for a hold
time $T=0$ leads to a plot indistinguishable from figure 5 (a). In this 
case the velocity of the side peaks is $\pm h/{(md)}=2.96$
$\mu$m/ms. After 15
ms of free expansion each side peak moves 44.4 $\mu$m consistent with the
simulation of figure 5 (a). In figure 5(b) it is blurred at $t = 6$ ms 
in agreement with  \cite{ari3}. The interference pattern is completely 
destroyed at $t=10$ ms and again revived after a half cycle of oscillation
at $t=18$ ms in figures 5 (c) and (d), respectively, in agreement with
experiment.

The explanation for the phenomenon observed in figures 5 is as follows. As
the system oscillates in the harmonic potential upon the removal of the
optical-lattice potential, after every half cycle of oscillation the
system passes through the same initial state when the optical-lattice
potential is removed \cite{adhi1}. This initial state is a coherent state
on the optical lattice and after a free expansion leads to the
interference pattern. However, if the system is allowed to evolve in the
harmonic potential for some time after removing the optical-lattice
potential and then subject to a free expansion a completely different
scenario emerges. In general, the three expanding bright spots would now
be moving freely in ``wrong" directions so that they may interact and
collide with each other upon removal of the harmonic trap 
and lose their identity resulting in a loss of the
observed interference pattern in figures 5.  This defocussing effect in
the interference pattern would be absent for hold times equal to a
multiple of half period of oscillation and would be minimal for hold times
near these values consistent with figures 5. There has been a recent
attempt to study a similar  phenomenon in a different context using a
one-dimensional mean-field model \cite{model}. 

\section{Conclusion}

In conclusion, using the explicit numerical solution of the
axially-symmetric GP equation we have studied the expansion of a BEC
formed in a joint harmonic and optical-lattice traps when either trap is
removed. When the harmonic trap is removed the piece of condensate in a
particular site expands radially without moving to the next site. As a
result the whole condensate expands radially maintaining a fixed axial
size. The shape of the condensate and its radial and axial sizes during
expansion are found to evolve in agreement with the experiments of
\cite{ari1,ari2}. When the optical trap is removed the pieces of the BEC
at different sites expand and interfere to form a distinct pattern with
three peaks which keep on oscillating according to (\ref{shm}) in
agreement with present simulation and experiment of \cite{ari3}. If the
interference pattern is allowed to evolve in the harmonic trap for some
hold time and the harmonic trap removed, prominent interference pattern
appears
for small hold time, which become blurred and finally, disappear  with the
increase of hold time in agreement with the experiment of
\cite{ari3}. Finally, for hold times larger than half period of
oscillation the interference pattern reappears as in experiment.

\ack  
We thank Dr. Oliver Morsch for very helpful correspondences on their
experiments \cite{ari1,ari2,ari3}. 
The work was supported in part by the Conselho Nacional de Desenvolvimento
Cient\'ifico e Tecnol\'ogico (CNPq) 
of Brazil.


\section*{Reference}
 
\begin{thebibliography}{99}
 
 
 
\bibitem{1}   Anderson B P and  Kasevich M A 1998 {\it Science} {\bf 282}
1686
 
 
 
\bibitem{2} Orzel C,
Tuchman A K,  Fenselau M L,  Yasuda M and  Kasevich M 2001
{\it Science} {\bf 291} 2386

 
\bibitem{greiner} Greiner M,  Mandel O,  Esslinger T,  H\"ansch T W
and  Bloch I  2002 {\it Nature (London)} {\bf 415} 39
 
 Greiner M,  Mandel O,   H\"ansch T W
and  Bloch I  2002 {\it Nature (London)} {\bf 419} 51

\bibitem{cata} Cataliotti F S,  Burger S,  Fort C,  Maddaloni P,
 Minardi F,
Trombettoni A,  Smerzi A and   Inguscio M 2001
{\it Science}
{\bf 293} 843

\bibitem{ady}
Adhikari S K 2003 {\it Eur. Phys. J. D} {\bf 25} 161

\bibitem{ari0} Morsch O,  M\"uller J H,  Cristiani M,   Ciampini D   and
 Arimondo E 2001 {\PRL}  {\bf 87} 140402 


 
 
\bibitem{cata2}  Cataliotti F S,  Fallani L,  Ferlaino F,  Fort C,
 Maddaloni P and   Inguscio M 2003 \NJP {\bf 5} 71


 
\bibitem{sol}
Strecker K E, Partridge G B, Truscott A G and Hulet R G 2002
{\it  Nature (London)}  {\bf 417} 150
 
 


 
\bibitem{th}  
  Smerzi A, Trombettoni A, Kevrekidis P G and
 Bishop A R 2002 {\it Phys. Rev. Lett } {\bf 89} 170402

Adhikari S K 2003  {\PL} A {\bf 308} 302


Adhikari S K 2003 \jpb {\bf 36} 2725

 Adhikari S K 2003  {\it Eur. Phys. J} D {\bf 25} 161
 


\bibitem{ari1}  Morsch O,  Cristiani M,  M\"uller J H,   Ciampini D
and 
 Arimondo E  2002 {\it
Phys. Rev. A} {\bf 66} 021601 

\bibitem{ari2}  Morsch O,  Cristiani M,  M\"uller J H,   Ciampini D
and
 Arimondo E 2003 {\it Laser Phys.} {\bf 13} 594


\bibitem{ari3}  M\"uller J H,    Morsch O,    Cristiani M,   Ciampini D
and
 Arimondo E  2003 \JOB {\bf 5} S38



 
 


\bibitem{8}  Dalfovo F,  Giorgini S,  Pitaevskii L P and
 Stringari S 1999 {\it
Rev. Mod.  Phys.} {\bf 71} 463





\bibitem{fexp} Adhikari S K 2002 \PR A {\bf  65}  033616 

Dalfovo F and Modugno M 2000 \PR A {\bf 61}  023605

Holland M and Cooper J 1996 \PR A {\bf 53} R1954

 Hosten O, Vignolo P, Minguzzi A, Tanatar B and   Tosi M P
2003 \jpb  {\bf 36 }  2455
 

 







 


\bibitem{11}Adhikari S K 2002  {\it Phys. Rev. E} {\bf 65}   016703


\bibitem{11x} Adhikari S K and  Muruganandam P 2002 {\it
J. Phys. B: At. Mol. Opt.} {\bf 35} 2831


  Muruganandam P and Adhikari S K 2003 \jpb {\bf 36} 2501


\bibitem{adhi1} Adhikari S K and  Muruganandam P  2003 {\PL} A {\bf  310}
229



 
\bibitem{xiong} Xiong H, Liu S, Huang G and Xu Z 2002 \jpb  {\bf 35} 4863

\bibitem{model}Morsch O, M\"uller J H,  Ciampini D,  Cristiani M, Blakie P
B, Williams C J, Julienne P S and Arimondo E 2003 
\PR A {\bf 67} 031603 
\end{thebibliography}
 \end{document}